\documentclass[twocolumn]{aastex63}
\usepackage{bm,amsmath,amssymb}
\usepackage{color,soul}

\def \FF {\bm{F}}
\def \ab {\alpha\beta}
\def \dela {\partial_{\alpha}}
\def \delb {\partial_{\beta}}
\def \nhat {\hat{n}}
\def \massp {m_{\rm p}}
\def \OmegaK {\Omega_{\rm K}}
\def \dmax {\delta_{\rm max}}
\def \dtyp {\delta_{\rm typ}}
\def \taum {\tau^{\rm min}_{\rm f}}
\def \taut {\tau^{\rm typ}_{\rm f}}

\def \mnot {m_{0}}

\def \rhod {\rho_{\bullet}}
\def \UU  {\bm{U}}

\def \vK   {v_{\rm K}}
\def \tauf {\tau_{\rm f}}
\def \taus {\tau_{\rm s}}
\def \ker {\kappa_{\rm er}}
\def \tstar {t_{\ast}}
\def \mdot {\dot{m}}

\def \taus {\tau_{\rm s}}

\def \rhog {\rho_{\rm g}}

\def \Vzero {V_{\rm 0}}
\def \Lzero {L_{\rm 0}}
\def \Rey  {\mbox{Re}}
\def \Ma  {\mbox{Ma}}
\def \Md  {\mbox{Ma}_{\rm d}}
\def \xx   {\bm{x}}
\def \vv   {\bm{v}}
\def \uu   {\bm{u}}
\def \qq   {\bm{q}}
\def \ab   {\alpha\beta}
\def \cab  {c_{\alpha}c_{\beta}}
\def \ca    {c_{\alpha}}
\def \grad {\bm{\nabla}}

\def \Er    {\mbox{Er}}
\def \Th    {\mbox{Th}}
\def \mnot {m_{\rm 0}}
\def \cs {c_{\rm s}}

\newcommand{\ATT}[1]{{\bf #1} }

\newcommand{\eq}[1]{~(\ref{#1})}
\newcommand{\Eq}[1]{Eq.~(\ref{#1})}
\newcommand{\Fig}[1]{Fig.~(\ref{#1})}
\newcommand{\subfig}[2]{Fig.~(\ref{#1}#2)}

\usepackage{tikz}
\usetikzlibrary{arrows.meta}
\usetikzlibrary{shapes, shadows, arrows}
\tikzstyle{block} = [draw, rectangle, fill=gray!30, text width=20em, text centered, minimum height=10mm, node distance = 15mm]
\tikzstyle{line} = [draw, -stealth, thick]

\newcommand{\myGlobalTransformation}[2]
{
    \pgftransformcm{1}{0}{0.3}{0.3}{\pgfpoint{#1cm}{#2cm}}
}

\tikzstyle myBG=[line width=3pt,opacity=1.0]

\newcommand{\drawLinewithBG}[2]
{
    \draw[black, opacity = 0.4] (#1) -- (#2);
}

\newcommand{\graphLinesHorizontal}
{
    \drawLinewithBG{1,1}{5,1};
    \drawLinewithBG{1,3}{5,3};
    \drawLinewithBG{1,5}{5,5};
}

\newcommand{\graphLinesVertical}
{
    \pgftransformcm{0}{1}{1}{0}{\pgfpoint{0cm}{0cm}}
    \graphLinesHorizontal;
}

\newcommand{\graphThreeDnodes}[2]
{
    \begin{scope}
        \myGlobalTransformation{#1}{#2};
        \foreach \x in {1,3,5} {
            \foreach \y in {1,3,5} {
                \node at (\x,\y) [circle,fill=gray,minimum size = 0.1cm, inner sep = 0cm] {};
            }
        }
    \end{scope}
}

\newcommand{\numberNodes}
{
    \begin{scope}
        \myGlobalTransformation{0}{0}
        \node (20) at (1,1) [below right]{20};
        \node (8) at (3,1) [below right]{8};
        \node (19) at (5,1) [below right]{19};
        \node (9) at (1,3) [below right]{9};
        \node (1) at (3,3) [below right]{1};
        \node (7) at (5,3) [below right]{7};
        \node (21) at (1,5) [below right]{21};
        \node (10) at (3,5) [below right]{10};
        \node (22) at (5,5) [below right]{22};
    \end{scope}
    \begin{scope}
        \myGlobalTransformation{0}{2.125}
        \node (12) at (1,1) [below right]{12};
        \node (3) at (3,1) [below right]{3};
        \node (11) at (5,1) [below right]{11};
        \node (4) at (1,3) [below right]{4};
        \node (0) at (3,3) [below right]{0};
        \node (2) at (5,3) [below right]{2};
        \node (13) at (1,5) [below right]{13};
        \node (5) at (3,5) [below right]{5};
        \node (14) at (5,5) [below right]{14};
    \end{scope}
     \begin{scope}
        \myGlobalTransformation{0}{4.25}
        \node (24) at (1,1) [below right]{24};
        \node (15) at (3,1) [below right]{15};
        \node (23) at (5,1) [below right]{23};
        \node (16) at (1,3) [below right]{16};
        \node (6) at (3,3) [below right]{6};
        \node (18) at (5,3) [below right]{18};
        \node (25) at (1,5) [below right]{25};
        \node (17) at (3,5) [below right]{17};
        \node (26) at (5,5) [below right]{26};
    \end{scope}
}

\newcommand{\farrows}[2]
{
    \begin{scope}
        \myGlobalTransformation{#1}{#2};
        \foreach \x in {1,3,5} {
            \foreach \y in {1,3,5} {
                \node (thisNode) at (\x,\y) {};
                {
                    \pgftransformreset
                    \draw[->, red] (3.9 , 3.1) -- (thisNode) node[above right] {};
                }
            }
        }
    \end{scope}
}

\graphicspath{{./}{figv2/}}
\begin{document}
\title{Planetesimals on eccentric orbits erode rapidly}
\correspondingauthor{Dhrubaditya Mitra}
\email{dhruba.mitra@gmail.com}
\author{Lukas Cedenblad}
\affiliation{NORDITA, Royal Institute of Technology and Stockholm University,
  Roslagstullsbacken 23, SE-10691 Stockholm, Sweden}
\affiliation{
    Department of Physics,
  AlbaNova University Center,
  Stockholm University, SE-10691
  Stockholm, Sweden.}
\author{Noemi Schaffer}
\affiliation{
  Lund Observatory,
  Department of Astronomy and Theoretical Physics,
  Lund University, Box 43, 22100 Lund, Sweden.}
\author{Anders Johansen}
\affiliation{
  Lund Observatory,
  Department of Astronomy and Theoretical Physics,
  Lund University, Box 43, 22100 Lund, Sweden.}
\affiliation{Centre for Star and Planet Formation, Globe Institute,
  University  of Copenhagen, Oster Voldgade 5–7, 1350 Copenhagen, Denmark}
\author{B. Mehlig}
\affiliation{Department of Physics, Gothenburg University, SE-41296 Gothenburg, Sweden}
\author[0000-0003-4861-8152]{Dhrubaditya Mitra}
\affiliation{NORDITA, Royal Institute of Technology and Stockholm University,
Roslagstullsbacken 23, SE-10691 Stockholm, Sweden}
\shorttitle{Planetesimals on eccentric orbits erode rapidly}
\shortauthors{L. Cedenblad et al}
\preprint{NORDITA 2021-084}
\begin{abstract}
  We investigate the possibility of erosion of planetesimals in
  a protoplanetary disk. 
  We use theory and direct numerical simulations (Lattice Boltzmann Method)
  to calculate the erosion of large -- much larger than the mean-free-path
  of gas molecules -- bodies of different shapes in flows. We find
  that erosion follows a universal power-law in time, at intermediate times,
  independent of the Reynolds number of the flow and the initial shape of the
  body. Consequently, we estimate that  planetesimals in eccentric orbits,
  of even very small eccentricity, rapidly (in about hundred years) erodes
  away if the semi-major axis of their orbit lies in the inner
  disk -- less than about $10$ au. 
Even planetesimals in circular orbits erode away in approximately ten thousand
years if the semi-major axis of their orbits are $\lessapprox 0.6$au.
\end{abstract}
%
\section{Introduction}
According to our present understanding, the
process of formation of planets 
begins with the growth of micrometer sized dust in a protoplanetary disk~\citep{Arm10}.
Dust particles  move around the central star in Keplerian orbits and at the
same time settle down to the midplane of the disk.
Let us assume that whenever two dust particles
collide, they stick together. 
Consider an aggregrate of dust particles -- a planetesimal
--  rotating around the central star. 
The planetesimal rotates with Keplerian speed while
the gas around it rotates at a slightly sub-Keplerian speed.
Hence, the planetesimal feels a \textit{headwind} thereby
loses angular momentum due to drag forces and spirals
into the star very rapidly in about a few hundred years.
Within this time the planetesimal is estamated to 
at most grow to the size of few meters~\citep{Youdin10,Arm10}.
This \textit{meter-sized barrier} appears to prevent
planet formation. 
Over the years, several possible solutions, including the streaming
instability~\citep{you+god05,johansen2007rapid},
concentration in vortices and pressure bumps~\citep{barge1995did,
  klahr2006formation, johansen2009zonal}
enhanced rate of collision due to turbulence~\citep{mitra2013can},
gravitational collapse of clouds of
pebbles~\citep{klahr2020turbulence, klahr2021testing},
to name a few, have been suggested, see also \cite{johansen2014multifaceted},
for a review.
To make matters worse, it is quite unlikely that whenever two dust particles
collide they stick.
Possible outcomes of collisions could be sticking,
complete or partial, fragmentation, or bouncing depending on the
mechanical, e.g., relative velocities on collision, and
thermodynamic (e.g., ambient temperature)
conditions~\citep{wilkinson2008stokes,blu+wur08,wet10,Zsom11}.
The dust aggregate that forms in this manner is likely to be very loosely
bound, therefore could it not break up ?
There are several possibilities, e.g., :
(a) two dust aggregates may collide and fragment
and
(b) the gas can erode the dust aggregate away.
In this paper, we investigate the second possibility which has received 
considerable attention recently~\citep{paraskov2006eolian,
  musiolik2018saltation,
  demirci2019pebble, kruss2020wind, schaffer2020erosion, rozner2020aeolian,
  demirci2020planetesimals}.

The rest of the paper is organized in the following manner.
We first consider the mathematical problem of erosion of a
solid by a fluid.
In section~\ref{sec:model} we describe our model
in terms of several dimensionless parameters
and next, following
\cite{ristroph2012sculpting,moore2013self} and \cite{huang2015shape},
present a theoretical framework to understand
this problem.
In section~\ref{sec:results} we show the results of our
numerical simulation of erosion using the Lattice Boltzmann Method~(LBM). 
In particular, we demonstrate that if the stress holding the body is
small enough, the solid erodes away in a finite time.
In section~\ref{sec:disk} we show the relevance of these results
for planetesimals in protoplanetary disks. 
We conclude in section~\ref{sec:conc}.
\section{Model}
\label{sec:model}
\begin{figure}
  \includegraphics[width=0.9\columnwidth]{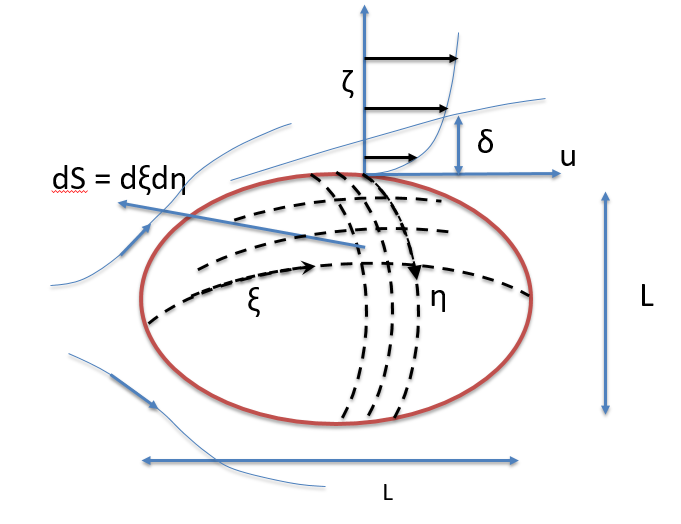}
  \caption{A sketch of a boulder of length scale $L$ with a flow around it.
    Far away from the body the velocity of the flow is $\UU$.
We use a set of a generalized coordinate system $\xi,\eta,\zeta$ such that
$\xi$--$\eta$ form a set of surface coordinates and $\zeta$ is the coordinate
perpendicular to the surface.
An infinitesimal surface area
$dS = d\xi d\eta$.
Ahead of the body is a point in the flow where the
velocity is zero relative to the boulder. This point is called the stagnation point.
We choose $\xi$ such that all the points with a constant $\xi$ are equidistant
from the stagnation point and as $\xi$ increases we move away from the stagnation point. 
The flow velocity is zero at the surface of the
boulder -- no-slip boundary condition.
The tangential component of velocity $u$ rises sharply
from zero to its free-stream value $\mid \UU \mid$ within a small distance $\delta$,
the thickness of the boundary layer. In the coordinate system chosen $\delta$ is a function
of $\xi$ alone. 
  }
  \label{fig:sketch}
\end{figure}
We consider a large boulder with a characteristic length scale $L$ moving with a
speed $U$ in a fluid with dynamic viscosity $\mu = \rho\nu$, where
$\rho$ is its density and $\nu$ its kinematic viscosity. 
We assume that the Reynolds number,
$\Rey \equiv UL/\nu$,
of the flow is large, but the flow is not otherwise
turbulent. 
Let us consider this problem in a frame fixed with the boulder.
In this frame the speed of the flow far away from the boulder is $U$.
We consider the following model of erosion~\citep{jager2017channelization,schaffer2020erosion}
:
the rate of mass--loss from an infinitesimal surface area $dS$ 
of the body is given by
\begin{equation}
  \mdot {\rm d}S=
  \begin{cases}
      \ker\left( \tauf -\taus \right) {\rm d}S \/, & \text{for}\quad \tauf > \taus \\
      0\/, &  \text{otherwise}\/. 
  \end{cases}
  \label{eq:erosion}
\end{equation}
Here, $\tauf$ is  the normal component of the
shear stress due to the fluid, $\taus$ is a threshold stress --
the solid starts eroding once the fluid stress exceeds this threshold stress --
and $\ker$ is a constant of proportionality.
Equation~\ref{eq:erosion} is purely empirical.
It is often used in estimations of erosion of river beds~\citep{Shields1936,Dey14}.
However, if we assume that the erosion rate is an analytic function of
$\tau \equiv \tauf - \taus$, then for small $\tau$ \Eq{eq:erosion} holds.
Both the erosion coefficient, $\ker$, and the threshold stress, $\taus$
depend on the material
properties of the solid, e.g., its composition and porosity.
Note that in \Eq{eq:erosion} $m$ has the dimension of
$\text{mass}/\text{area}$ and
$\ker$ has the dimension of inverse velocity.
Let us choose a set of generalized coordinates $\xi,\eta,\zeta$ such that
$\xi$--$\eta$ form a set of surface coordinate
and $\zeta$ is the coordinate
perpendicular to the surface, see \Fig{fig:sketch}.
The fluid stress $\tauf \equiv \mu \partial u/\partial\zeta $
where $u$ is the component of the flow velocity along the tangential direction.
If the fluid stress is larger than the solid stress, the boulder starts eroding.
We also assume that the erosion proceeds on a characteristic time scale much
slower than $T \equiv L/U$.
As the body erodes, the flow around the body changes, this in turn changes
the fluid stress and hence the rate of erosion.
This is an example of a free boundary problem.
Ahead of the body is a point in the flow whose
velocity is zero relative to the boulder. This point is called the stagnation point.
We choose $\xi$ such that all the points with a constant $\xi$ are equidistant
from the stagnation point and as $\xi$ increases we move away from the stagnation point\footnote{For example,  if the boulder is a sphere of radius $L$ we choose a spherical
  polar coordinate systems with the $z$--axis pointing along the flow.
  Then, $d\xi = Ld\theta$ and $d\eta = L \sin(\theta)d\phi$,
  and $dS = L^2\sin(\theta)d\theta d\phi$ where
  $\theta$ is the polar angle and $\phi$ the azimuthal angle, respectively.
  Lines of constant $\xi$ are the latitudes of this sphere and the lines
  of constant $\eta$ the longitudes. The stagnation point lies outside
  the sphere somewhere on the $z$--axis. 
}. 
The flow velocity is zero at the surface of the
boulder -- no-slip boundary condition. The tangential component of velocity,
$u$, rises sharply
from zero to its free-stream value $U$ within a small distance $\delta$,
the thickness of the boundary layer.
The theory of laminar boundary layer~\citep[see, e.g.,][chapter IV]{LLfluid}
estimates $\delta$ as
\begin{equation}
  \delta = \sqrt{\frac{\nu \xi}{U}}\/,
  \label{eq:delta}
\end{equation}
valid for $\xi$ not too close to zero, i.e., away from the stagnation point.
This allows us to estimate the fluid stress as
\begin{equation}
  \tauf \sim \mu \frac{U}{\delta}\/.
  \label{eq:tauf}
\end{equation}
\subsection{Dimensionless numbers}
We use $L$ as our characteristic length scale and $U$ as our characteristic velocity
scale to obtain $T \equiv L/U$ as our characteristic timescale.
We define the erosion number to be $\Er \equiv \ker U$. 
If the typical fluid stress is larger than the critical solid stress,
we expect erosion.
We define a corresponding dimensionless number, the threshold number ($\Th$)
\begin{equation}
  \Th \equiv \frac{\tauf}{\taus}\/.
  \label{eq:th}
\end{equation}
Erosion happens only if $\Th > 1$. 
The three dimensionless numbers that completely specify our problem are:
the Reynolds number ($\Rey \equiv UL/\nu$), the erosion number ($\Er$) and
the dimensionless threshold~($\Th$).
To obtain a typical value for the Erosion number and the dimensionless
threshold,
we need to estimate a typical value for the fluid stress. 
We use two different estimates for the thickness of the boundary layer, and
consequently two different estimates for the fluid stress:
\begin{subequations}
  \begin{align}
  \dmax = \sqrt{\frac{\nu L}{U}} &\quad \dtyp = \sqrt{\frac{\nu \Lambda^2 \lambda}{U}} \\
  \taum = \frac{\rho U^2}{\sqrt{\Rey}} &\quad \taut = \frac{\rho U^2}{\Lambda\sqrt{\Ma}} \\
  \Th_{\rm min} = \frac{1}{\sqrt{\Rey}}\frac{\rho U^2}{\taus} &\quad
  \Th_{\rm typ} = \frac{1}{\Lambda\sqrt{\Ma}}\frac{\rho U^2}{\taus} 
\end{align}
\label{eq:param}
\end{subequations}
In the left column of \eq{eq:param}, we have used the length of the eroding body
$L$ as the length scale that determines the maximum value of the boundary
layer thickness which
corresponds to minimum value of the fluid stress.
In the right column, to estimate the typical value of the fluid stress we use
a length scale that is $\Lambda^2$ times the mean--free--path $\lambda$,
where $\Lambda^2$, the inverse Knudsen number, is large -- about $100$.
We also use   $\nu \sim \cs\lambda$ where $\cs$ is the speed of sound
-- a familiar result from the kinetic theory of gases~\citep[see, e.g.,][]{LLphykin}.
\subsection{Theoretical framework}
\label{sec:theory}
Recently, a collection of remarkable
papers~\citep{ristroph2012sculpting,moore2013self,huang2015shape} 
studied erosion of bodies in fluid flows both analytically and experimentally. 
For the sake of completeness we summarize their arguments
below.

First, assume that $\taus$ is so small that it can be safely ignored.
We can then estimate the rate of total mass loss as
\begin{eqnarray}
  \frac{dM}{dt} &\approx& -\ker\int\frac{\mu U}{\delta}d\xi d\eta 
  = -\ker\frac{\mu U^{3/2}}{\sqrt{\nu}}\int \frac{d\xi d\eta}{\sqrt{\xi}} \\
    &=& - \rhog\ker \sqrt{U^3\nu}L^{3/2} \/.
\label{eq:mloss}
\end{eqnarray}
Here, $M(t)$ is the total mass of the body. 
Next, assume the material density of the body to be a constant, $\rhod$.
Then, \Eq{eq:mloss} can be written as a differential equation for the instantaneous
volume $V(t)$,
\begin{equation}
  \frac{dV}{dt} = -C\ker\left(\frac{\rhog}{\rhod}\right)\sqrt{U^3\nu}\sqrt{V}\/.
  \label{eq:mloss3}
\end{equation}
We integrate this differential equation, with the initial condition that at $t=0$ the volume was
$\Vzero$, to obtain
\begin{equation}
  \frac{V}{\Vzero} = \left(1 - \frac{t}{\tstar}\right)^2\/,
  \label{eq:mloss4}
\end{equation}
with
\begin{equation}
   \frac{\tstar}{T} = \left(\frac{\rhod}{\rhog}\right)\frac{1}{C\Er} \sqrt{\Rey} \/,
\label{eq:tstar}
\end{equation}
where $\Lzero\equiv \Vzero^{1/3}$, and $\Rey \equiv U\Lzero/\nu$ is the
Reynolds number of the body at its initial size.
Given an initial volume, $\Vzero$, the characteristic time by which it erodes away
is $\tstar$ given in \Eq{eq:tstar}.
The constant $C$ is a constant that depends on
the shape of the body.

It is important to emphasize here that this theory shows that the process of
erosion is a power-law in time; hence, we cannot meaningfully define a characteristic
time scale of erosion or a rate of erosion.
The only meaningful time scale is the time scale $\tstar$. 

Several simplifying assumptions made above must now be qualified.
First, the expression for the boundary layer is for a laminar boundary layer,
strictly speaking, valid for small Reynolds number and also if the body is not
too large.
As the Reynolds number of the
flow increases, the boundary layer separates~\citep{LLfluid}; hence,
upper limit of the the integral over $\xi$
is not the dimension of the body, $L$, but a fraction of it.
The fraction itself is Reynolds number dependent -- decreases with Reynolds
number.
Thus we expect that for large Reynolds number the $\tstar$ in reality is larger
than the one obtained in \Eq{eq:tstar}.
There is a second, crucial, implicit, assumption of scale invariance in
deriving \Eq{eq:tstar}:
the shape of the body does not change as the body erodes.
This assumption is used in two places, once while assuming that the constant $C$
does not depend on time and a second time while assuming that there is only one,
time-dependent, length-scale $L$ that determines the time-dependent volume.
In other words, the body erodes in a self-similar manner.
This may not be true at the initial stages of erosion -- erosion at initial times may
depend on the initial shape of the body -- hence, the power-law dependence of
volume on time may not be observed in early stages of erosion. 
Finally, note that while arriving at \Eq{eq:mloss} we have assumed that $\xi$
and $\eta$ can be
integrated independent of each other; this assumption can be relaxed to obtain
essentially the same result. 

The equations~\ref{eq:mloss4} and \ref{eq:tstar}
are essentially a reworking of the results elucidated
by~\cite{ristroph2012sculpting}  and \cite{moore2013self}, 
who instead of writing an equation for evolution of  volume wrote one for the
surface area
which was confirmed by their experiments.

\subsection{Direct numerical simulation}
We study erosion by direct numerical simulation.
This poses a difficult problem because we have to be able to solve the equations
of the flow with an irregular boundary which itself evolves with time.
Most Navier--Stokes solvers are unable to deal with such a problem.
We choose to use the Lattice Boltzmann Method (LBM).

The Lattice Boltzmann Method, which is a descendant of the lattice gas
algorithm, is used quite commonly in fluid mechanics.
Hence, we do not give a detailed description of the algorithm here.
It is described in great detail in several reviews~\citep{chen1998lattice,
  benzi1992lattice}
and books~\citep{Sukop2007,Succi2018lattice}. 
However, as its use in astrophysics is not very common, we do provide a short
description in Appendix~\ref{sec:LBM}.
We follow ~\cite{jager2017channelization} to implement erosion in our code,
see section \ref{sec:dns_erosion} for further detail.

Here, it is sufficient to mention a few important aspects of our simulations.
At the start of the simulation, grid points are classified as one of the three
types: solid, fluid, and interface. 
In contrast to the theoretical framework presented in section~\ref{sec:theory}
we do implement a threshold value for the solid stress.
If the fluid stress exceeds this threshold value, an interface point loses
mass following the empirical law of erosion, \Eq{eq:erosion}. 
Once the loss of mass exceeds a certain fixed value, $\mnot$, an interface point
is changed to a fluid point and its erstwhile solid neighbors turn into interface.
This introduces a new dimensionless parameter $\mnot/(\rhod L^3)$.
In reality, erosion is not a continuous process in time --
it happens through sudden erosion of macroscopic dust grains.
The parameter $\mnot$ corresponds to the mass of such dust grains.

We benchmark our code, without the implementation of erosion, against
standard test cases.

\section{Results}
\label{sec:results}
\begin{figure*}
  \includegraphics[width=0.3\linewidth]{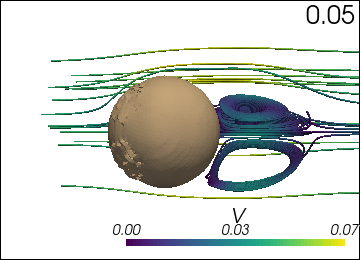}
  \includegraphics[width=0.3\linewidth]{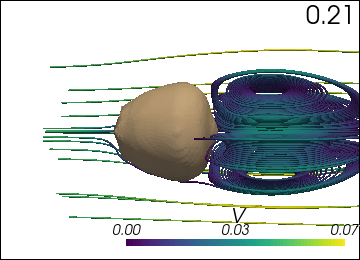}
  \includegraphics[width=0.3\linewidth]{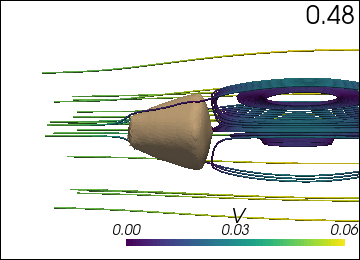}\\
  \includegraphics[width=0.3\linewidth]{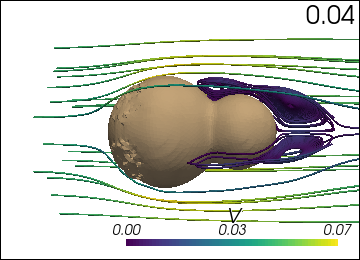}
  \includegraphics[width=0.3\linewidth]{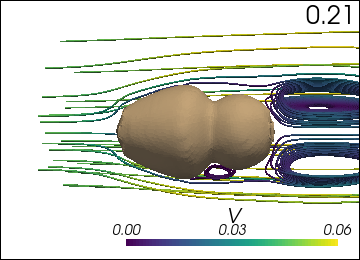}
  \includegraphics[width=0.3\linewidth]{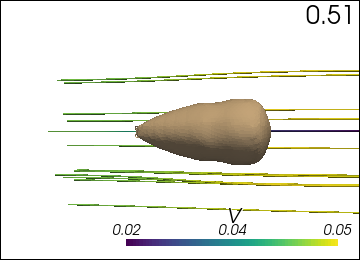}\\
  \includegraphics[width=0.3\linewidth]{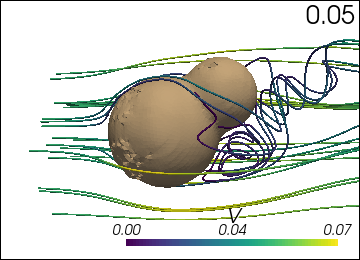}
  \includegraphics[width=0.3\linewidth]{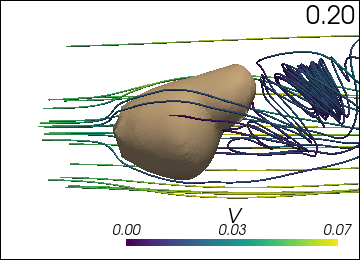}
  \includegraphics[width=0.3\linewidth]{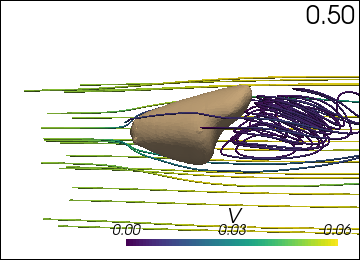}
  \caption{\label{fig:snowman} Erosion of  sphere and \textit{snowman} at
    $\Rey=800$ :
    the shape and topography of the sphere and the \textit{snowman} as a function of time.
    We also plot some of the streamlines of the flow around the eroding body.
    Time in units of $\tstar$ is shown at the top right corner. 
    Top row: the sphere.
    Middle row:  the \textit{symmetric snowman}, where the axis of the snowman is
    along the direction of the flow.
    In the bottom row, we plot the \textit{asymmetric snowman}, where the axis
    of the snowman makes an angle with the flow. The blue streamlines show
    the vortices. The location and strength of these vortices depend on
    the shape of the snowman and its direction relative to the flow.
    These vortices make a significant contribution to erosion but they
    are not part of the theoretical framework. The color bar shows
    magnitude of velocity measured in units of sound speed. The simulations
    are performed using Lattice Boltzmann Method with a resolution of
    $120\times 60\times 60$ grid points. }
\end{figure*}
 We first study the case of erosion of a
solid sphere followed by that of a
solid cube. 
Here, we present the results of the simulation of erosion of an irregular object
we call the \textit{snowman} which is made by merging two spheres -- one
smaller than the other, \Fig{fig:snowman}.
We do this for two important reasons: first, as our principal motivation is
erosion of planetesimals we consider a shape close to one that is
expected to be typical of planetesimals;
second, we expect that the departure from the theory, if any, is
larger for such irregular--shaped objects.

\subsection{Universal erosion}
\begin{figure}
  \includegraphics[width=0.9\linewidth]{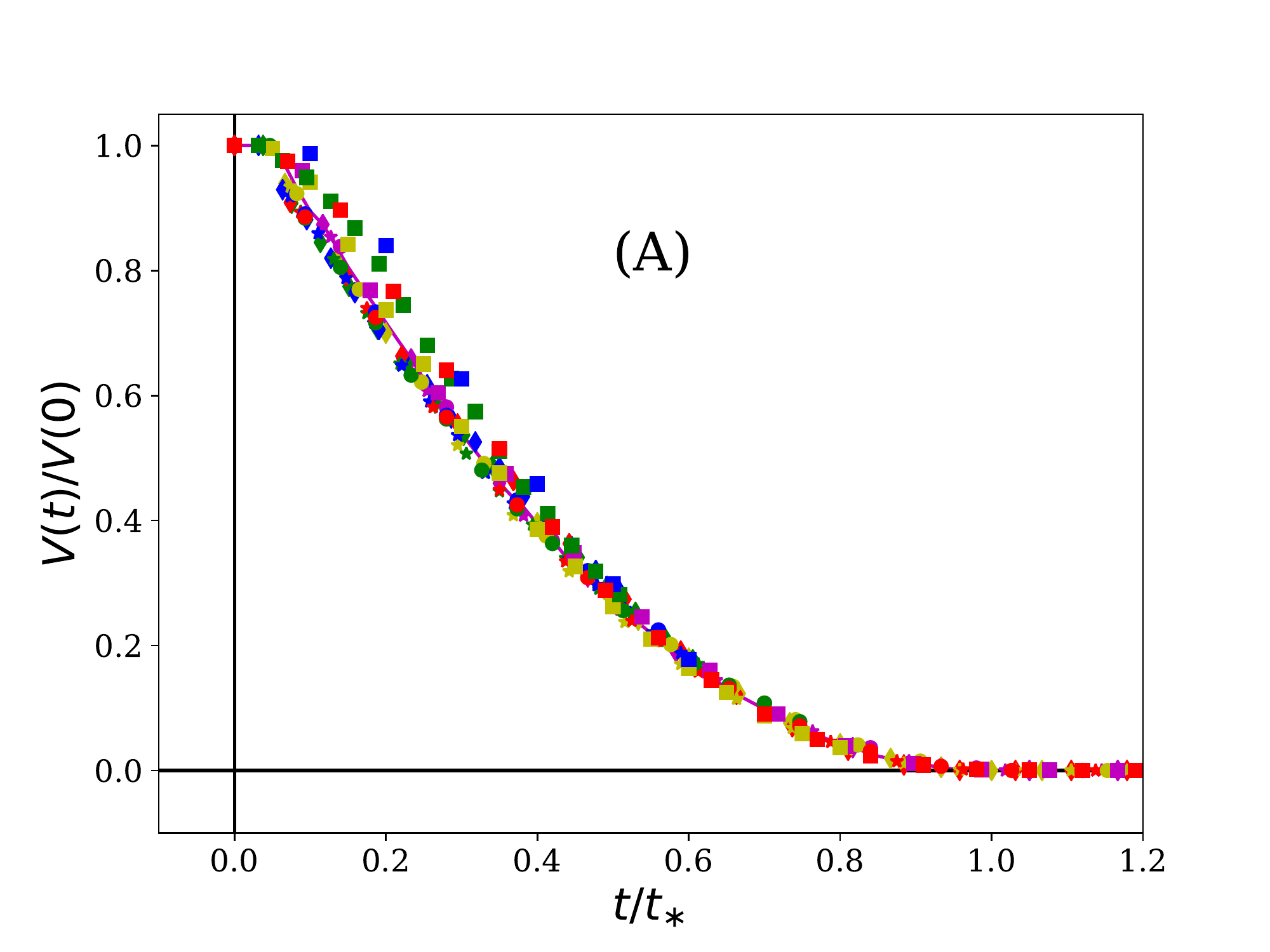}\\
  \includegraphics[width=0.9\linewidth]{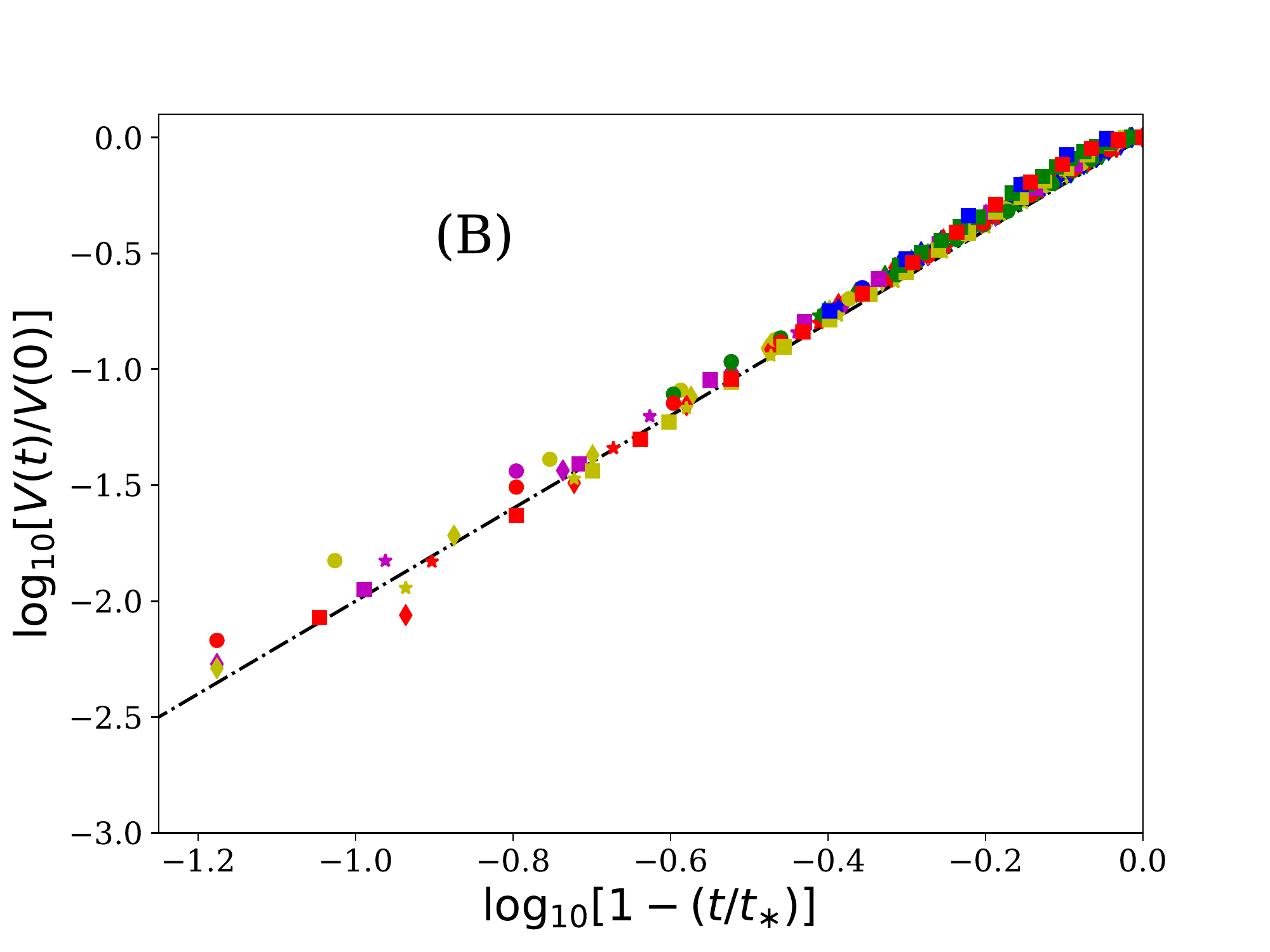}
  \caption{\label{fig:Vt} (A) Volume (normalized by the initial volume) of the
    solids as a function of time (normalized by $\tstar$) for all our runs.
    The colors label Reynolds number: $100$~(magenta), $200$~(yellow),
    $400$~(green), $~500$(blue), $800$~(red).
    The symbols label the shapes: $\diamond$~(snowman with axis along the flow, symmetric
    snowman), $\ast$~(snowman with axis not long the flow, asymmetric snowman),
    $\bullet$~(sphere) and $\square$~(cube). 
    (B) The same plot in log--log scale with the abscissa changed to,
    $1-t/\tstar$.
    The black dashed line has a slope of $2$. }
\end{figure}
\begin{figure}
  \includegraphics[width=0.9\columnwidth]{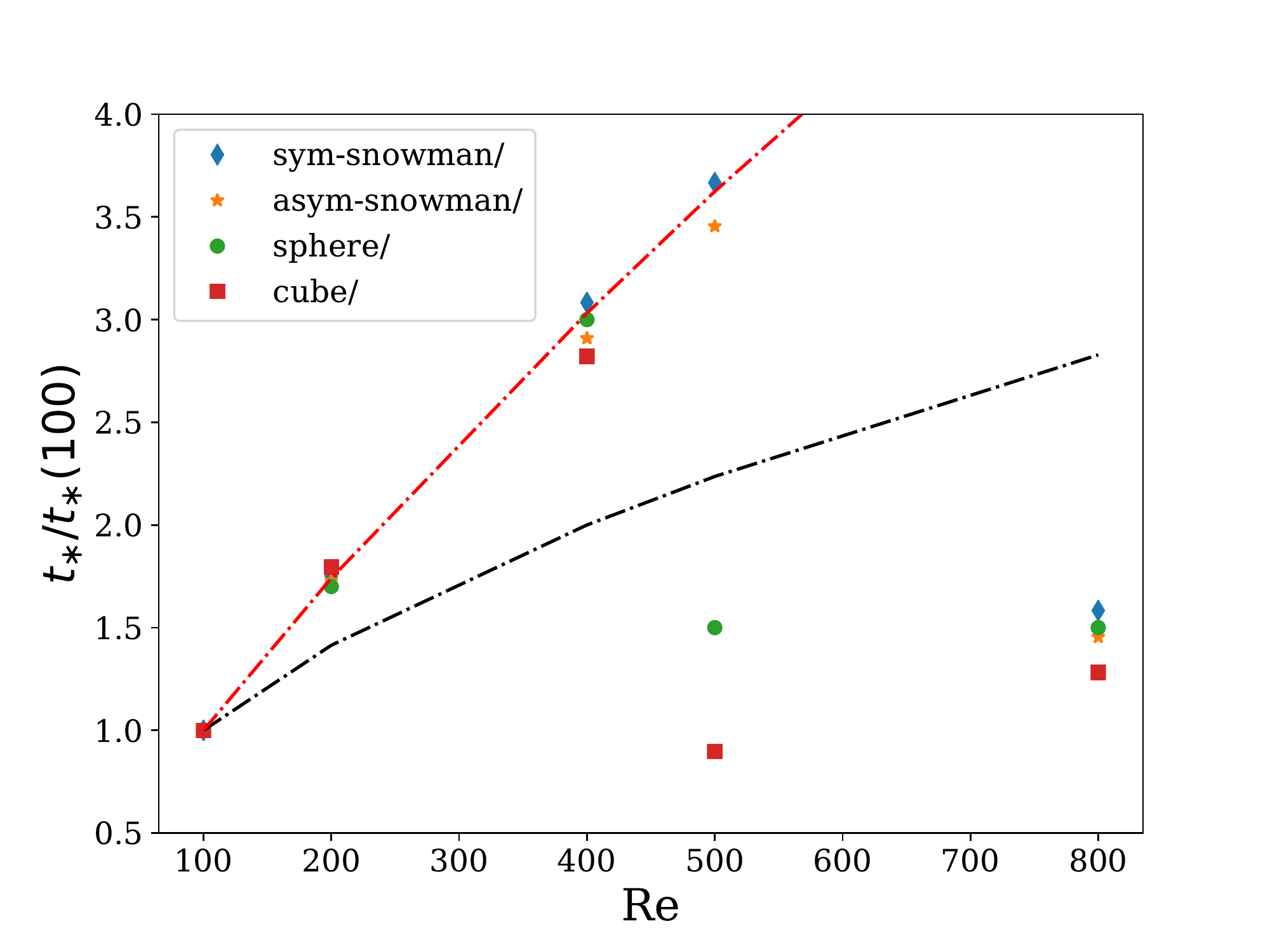}
  \caption{\label{fig:tstar_vs_Re}  Value of $\tstar$
    from our simulations
    as a function of the Reynolds number. The black dashed lines is $\sqrt{\Rey}$ and
    the red dashed line is $\Rey^{0.8}$.
    }
\end{figure}
In \Fig{fig:snowman} we show three different stages of erosion
for a snowman with its axis oriented along the flow (symmetric snowman)
and with an angle with the flow (asymmetric snowman) for $\Rey=800$.
In reality, the eroding body is expected to rotate, which
we ignore.
In \subfig{fig:Vt}{A} we show how the volume ($V(t)/V(0)$) of the snowman, the
sphere, and the cube changes as a function of time,
for five different Reynolds numbers:
$\Rey = 100, 200, 400, 500,$ and
$800$.
Remarkably, erosion for  all these solids, irrespective of the
Reynolds number, follows the
same \textit{universal law at intermediate times}.
There is departure from this law at early times,
particularly so for the cube, because erosion at early times
is not universal but depends on the shape of the eroding object. 
In \subfig{fig:Vt}{B}, we plot the volume
as a function of $1-t/\tstar$ in log-log scale. 
The theoretical expression, \Eq{eq:mloss3},
shown as a black dashed line, 
is a very good approximation to our numerical results,
except at late times.
We do expect this departure at late times because when the
solid becomes small the theory no longer applies.

\begin{figure}
  \includegraphics[width=0.7\columnwidth]{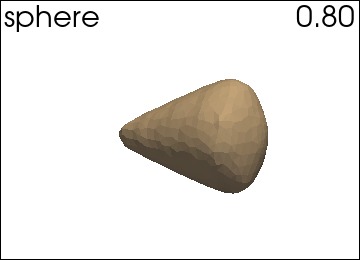} \\
  \includegraphics[width=0.7\columnwidth]{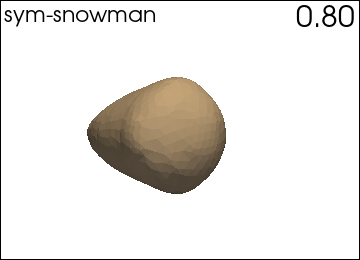} \\
  \includegraphics[width=0.7\columnwidth]{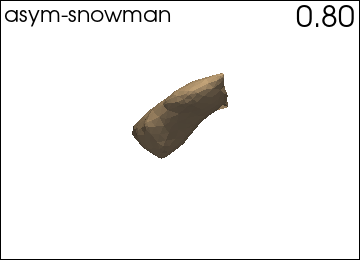}
  \caption{\label{fig:shape_late} The shape of the eroded body at 
	late times $t/\tstar \approx 0.8$. The flow, which is not
	shown here, is from left to right. 
    }
\end{figure}

Does the self-similar evolution of volume imply that
eroding bodies, irrespective of their initial shape, reduces
to the same shape?
We find that this is not the case.
In the last column of  \Fig{fig:snowman} 
we plot the shape of the eroded object for the three different 
initial shapes; the sphere, the symmetric snowman, and 
the asymmetric-snowman for $t/\tstar \approx 0.5$ 
-- a time at which the self-similar evolution holds. 
The three shapes are quite different from each other.
Even at very late times, $t/\tstar \approx 0.8$,
shown in \Fig{fig:shape_late}, the 
three different initial shapes do not become similar to each other.

Next, in \Fig{fig:tstar_vs_Re} we plot the time scale $\tstar$
as a function of $\Rey$ for different shapes.
The black dashed lines shows the expected dependence $\tstar \sim \sqrt{\Rey}$
-- clearly they do not agree.
For moderate $\Rey$ we find, $\tstar \sim \Rey^{0.8}$, shown as a red dashed line.
This disagreement is due to several reasons.
First, note that for any shape $\tstar$ is a non-monotonic function of $\Rey$.
This is because beyond a critical $\Rey$ the boundary layer 
separates~\cite[see, e.g.,][section 40]{LLfluid}.
This implies that the upper limit of the integral in \Eq{eq:mloss} over $\xi$
is no longer $L$ but smaller.
This suggest that the erosion time should increase (as rate of mass loss decreases)
but this is not the case in practice.
Instead, the separation of boundary layer is accompanied by
appearance of vortices behind the solid, we show several such examples
in \Fig{fig:snowman}. 
These vortices are very efficient at eroding the solid
thereby decreasing the erosion time by a large amount. 
But they are not accounted for in the theory we have described. 
Second, our numerical estimate of the erosion time, $\tstar$ is not very accurate.
We estimate it by recording the time the solid disappears but
the theory no longer applies as the solid becomes too small.

To summarize, our simulations show that the law of erosion given
in \Eq{eq:mloss4} holds, at intermediate times, for all the Reynolds numbers
and the shapes we have studied suggesting that \eq{eq:mloss4} is a
universal law of erosion. 
But the dependence of $\tstar$ on $\Rey$ is more complicated than the
simple expression: $\tstar \sim \sqrt{\Rey}$. 
\subsection{Erosion of planetesimals}
\label{sec:disk}
\begin{figure}
  \includegraphics[width=0.9\linewidth]{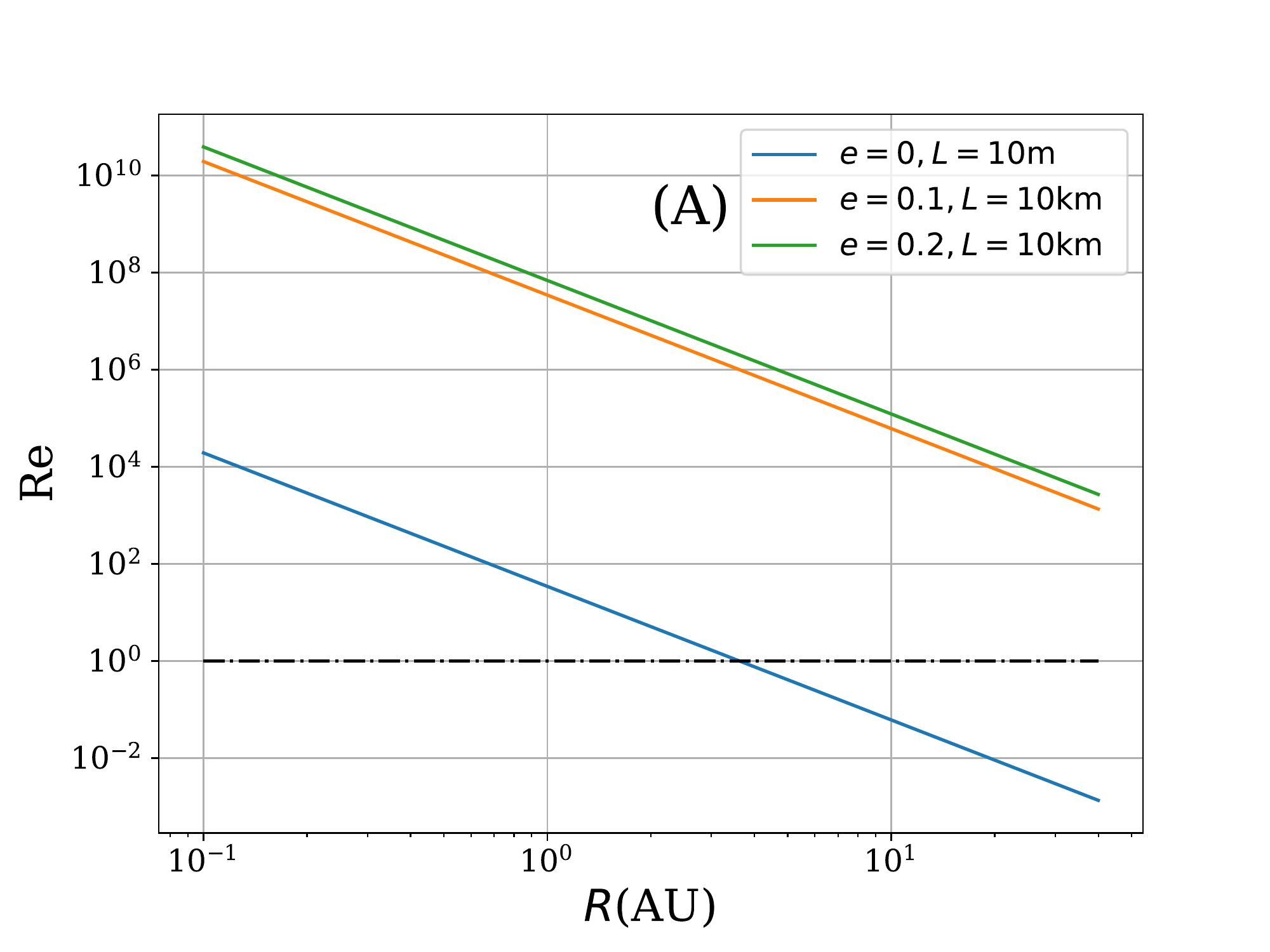} \\
  \includegraphics[width=0.9\linewidth]{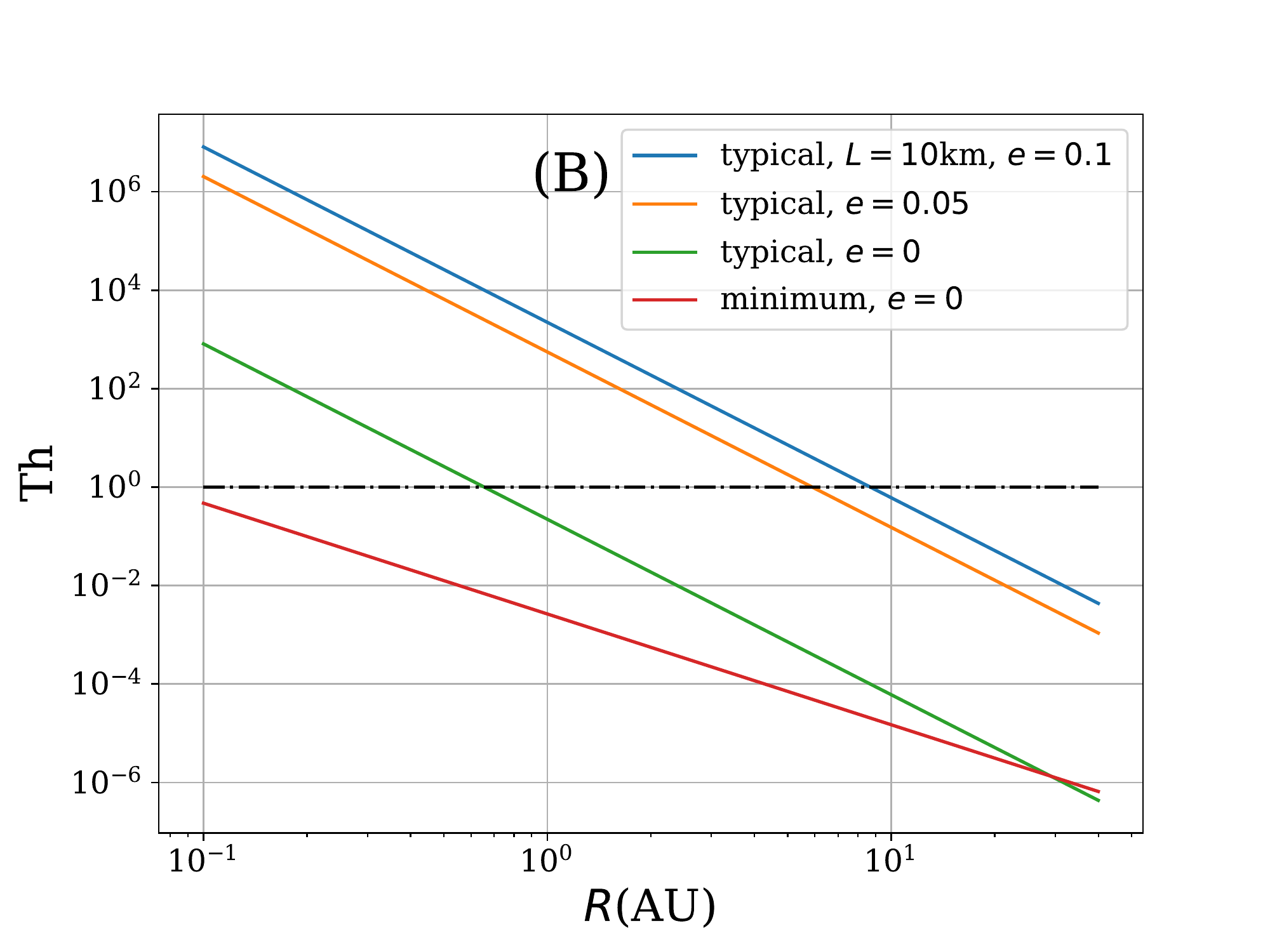}
  \includegraphics[width=0.9\linewidth]{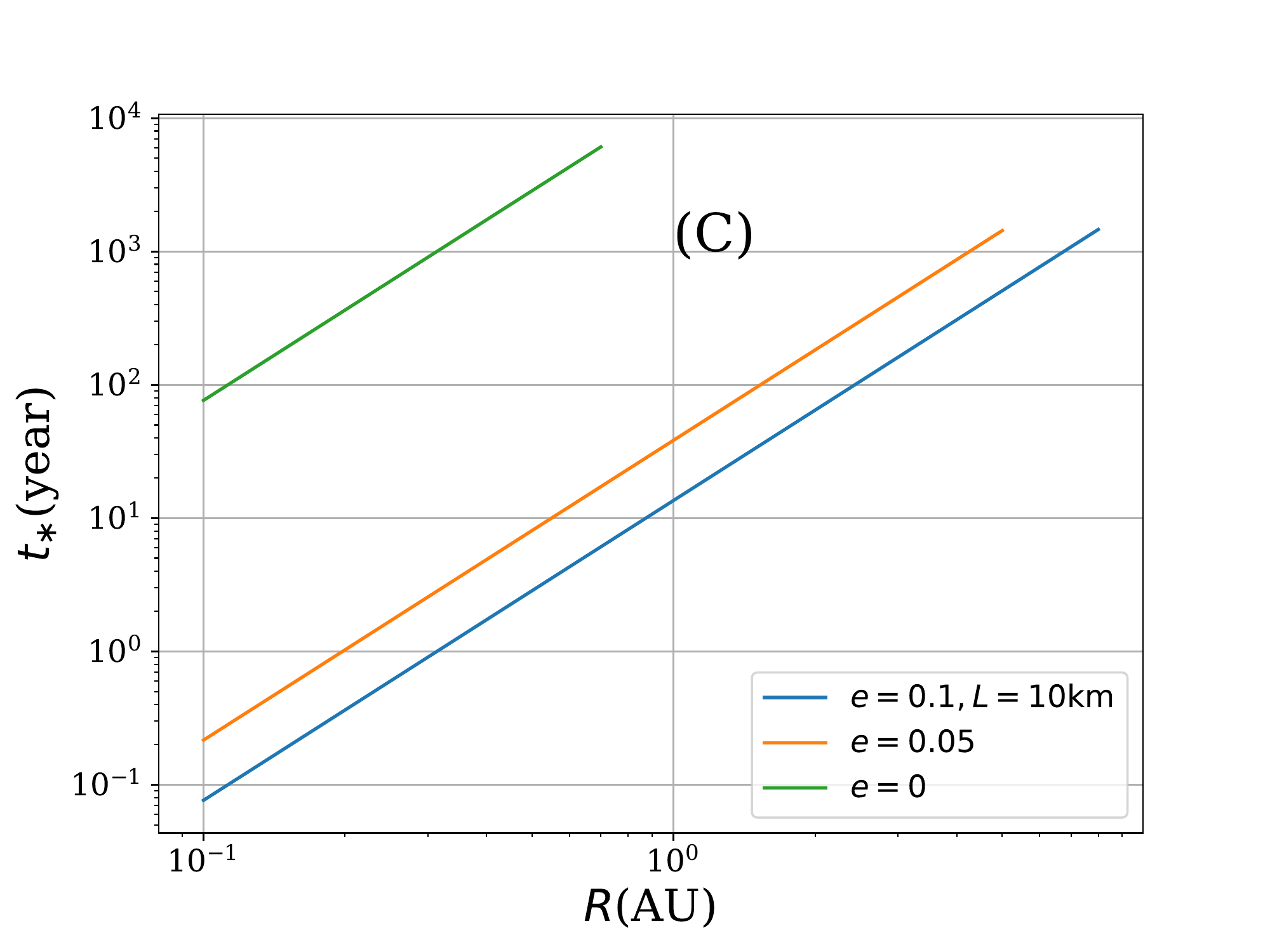}
  \caption{\label{fig:param}
  We consider a $10$ km sized body at a distance $R$
    from the central star with one solar mass. The exponents for the surface
    density and sound speed, see \Eq{eq:diskmodel}, are $\gamma=3/2$ and
    $\beta = 3/8$. We have also chosen $\Lambda = 10$.
    (A) The Reynolds number as a function of $R$ (in astronomical units) and 
  (B) The dimensionless threshold, $\Th$, as a function of $R$ (in astronomical units).
  The threshold stress $\taus = 0.0011$ Pa. 
  (C) The time it takes for a $10$km sized planetesimal to erode away
  as a function of the semi-major axis of its orbit calculated using
  \Eq{eq:tstar}.  
  We use $\ker \approx 150 {\rm m}^{-1}{\rm s}$ and
   $\rhod = 536{\rm kg}\/{\rm m}^{-3}$~\citep{patzold2019nucleus}.
   For each of the three cases we plot, 
  from  (B)  we calculate the maximum $R$ below which $\Th >1$,
  i.e. erosion occurs.  For planetesimals on orbits whose semi-major axis
  is greater than this maximum $R$ \Eq{eq:tstar} is no longer valid
  because such planetesimals do not erode.     
  }
\end{figure}
To understand the implication of our results for erosion of bodies
in protoplanetary
disks we need to estimate the dimensionless numbers for the disk.
We choose a simple model for protoplanetary disks:
\begin{equation}
  \Sigma \sim R^{-\gamma} \quad\text{and}\quad \cs \sim R^{-\beta}\/,
  \label{eq:diskmodel}
\end{equation}
where $R$ is the distance from the central star,
$\Sigma$ is the surface mass density and $\cs$ is the speed of sound. 
For example, in the minimum mass solar Nebula
model~\citep{hayashi1981formation, Arm10} $\gamma = 3/2$. 
The vertical scale height, $h(R)$, of the disk is $h \equiv \cs/\OmegaK$
where $\OmegaK \equiv \sqrt{GM_{\odot}/R^3}$ is the Keplerian velocity
($M_{\odot}$ is the mass of the central star and $G$ the gravitational
constant);
the density at the midplane is
$\rho(R) \sim \Sigma/h$;
the particle number density
$n(R) = \rho(R)/\massp$, where $\massp$ is the mass of proton,
the mean--free--path,
$\lambda(R) = 1/n\sigma$, where $\sigma \approx 2\times 10^{-15} {\rm cm}^2$
is the cross section of molecular collisions,
and the Mach number of the disk is
$\Md(R) \equiv \vK/\cs \sim R/h$. 
A planetesimal is typically in an orbit around
the central star with Keplerian speed
$\vK=\sqrt{GM_{\odot}/R}$.
The gas also rotates around the central star with a velocity close to but not
equal to the Keplerian velocity.
The difference in velocity is seen as a headwind by the
planetesimal~\citep{Arm10}.
The boulder in section~\ref{sec:theory} corresponds to a planetesimal and the 
velocity of this headwind corresponds to $U$. 
For a planetesimal on an eccentric orbit,  the headwind is not a constant but
depends on position of the planetesimal and the details of its orbit.
We consider an orbit with zero inclination.
At the position where the cosine of its
true anomaly is zero, the headwind is given by~\citep{adachi1976gas}
\begin{equation}
U=\vK\sqrt{e^2+\eta^2},
\label{eq:headwind}
\end{equation}
where $\eta \propto \cs^2/\vK^2$, is a dimensionless
number that can be as small as $10^{-3}$ depending on the position in the
disk.
Hence, we obtain $  T\OmegaK = T_{\rm K} (2\pi L/R)(1/\sqrt{e^2+\eta^2})$.
The Reynolds number of a solid of size $L$ in a orbit at a distance $R$
from the central star can be estimated to be~\citep[see, e.g.,][]{mitra2013can}
\begin{eqnarray}
  \Rey &=& \Ma\frac{L}{\lambda} \approx \sqrt{e^2+\eta^2}\Md\frac{L}{\lambda} \nonumber \\
  &\approx& \sqrt{e^2+\eta^2}\left(\frac{R}{h}\right)\left(\frac{L}{\lambda}\right) \/. 
\label{eq:Re}
\end{eqnarray}
The Mach number of the headwind is related to the disk Mach number,
$\Md \equiv \vK/\cs \approx R/h$, where $h$ is the vertical scale height of
of the disk at a radius $R$ from the central star,
$\Ma = \sqrt{e^2+\eta^2}\Md$.
Substituting these expressions in \Eq{eq:param}, we find how the
dimensionless parameters depend on the radial coordinate of the disk:
\begin{subequations}
  \begin{align}
    \Rey &\sim R^{2\beta-\gamma-2} \\
    \Er &\sim R^{-1/2} \\
    \Th_{\rm min} \sim R^{\gamma/2 -1} &\quad \Th_{\rm typ} \sim R^{\beta/2 - \gamma -9/4} 
  \end{align}
\end{subequations}

To give a specific example, we now consider a body with $L = 10$ km at a
distance $R$ from the central star with one solar mass.
The exponents for the surface density and sound speed, see \Eq{eq:diskmodel},
are $\gamma=3/2$ and $\beta = 3/8$ -- a minimum mass Solar Nebula model
for a razor-thin disk. 
In \subfig{fig:param}{A}, we plot $\Rey$ as a function of $R$. 
Estimation of the other two dimensionless numbers $\Er$ and $\Th$ is less
certain.

Let us first consider the dimensionless threshold, $\Th$. 
Experiments in laboratory~\citep{white1987saltation,paraskov2006eolian}
have tried to estimate the threshold stress necessary to lift dust grains 
from the surface of a pile of grains.
Unlike these experiments, erosion of a boulder in a protoplanetary disk 
does not depend on gravity.
Experiments in microgravity~\citep{musiolik2018saltation, demirci2019pebble,
  kruss2020wind, demirci2020planetesimals}
have tried to approach lower and lower gravity
and ambient pressure to get as close to the condition of protoplanetary disks
as possible.
The last of these~\citep{demirci2020planetesimals} measured the
critical shear stress
of a pile of glass beads in a parabolic flight campaign.
The critical shear stress depends on the size of the glass beads
and ambient pressure~\citep[ figure 5]{demirci2020planetesimals}.
They conclude that ``..cohesion is really low. At zero gravity,
the shear stress required to initiate erosion is only $0.0011$ Pa '',
i.e., $\taus \approx 0.0011$ Pa!
Substituting in \Eq{eq:param} we calculate both the typical value of $\Th$
and its minimum value.
In \subfig{fig:param}{B}, we plot the typical value of $\Th$
for orbits with three different eccentricities, $e=0.1, 0.05,$ and
zero (circular orbit). For the first one, the dimensionless threshold
remains greater than unity for $R \lesssim 9{\rm au}$.
For the second one, $e=0.05$, the dimensionless threshold
remains greater than unity for $R \lesssim 8{\rm au}$.
For orbits of even higher eccentricities erosion remains important
for even larger values of $R$.
For an orbit of zero eccentricity the typical value of $\Th$ remains
greater than unity for $R \lesssim 0.6{\rm au}$.
If instead of the typical value of $\Th$ we consider its minimum value
then the dimensionless threshold for a circular orbit is less than unity
everywhere. 
We conclude that typically, erosion occurs for eccentric orbits,
even with eccentricity as small as $0.05$ in the inner disk.
Erosion happens even for boulders in perfectly circular orbits
if they are close enough to the central star. This
  result is different from our earlier work~\citep{schaffer2020erosion}
  because of two reasons: (a) In the light of recent
  experimental results~\citep{demirci2020planetesimals} we consider
  a lower value of $\taus$. (b) We consider the typical fluid stress
  not the minimum value as we had done before.

There is even less experimental data to estimate the  erosion number $\Er$.
As in our earlier work~\citep{schaffer2020erosion},
following \cite{demirci2019pebble}, we assume a value of
$\ker \approx 150 ({\rm s}/{\rm m}$.
In \subfig{fig:param}{C} we plot the time it takes for body
to erode away, $\tstar$ from \Eq{eq:tstar}, with
$L = 10$km, and $\rhod = 536{\rm kg}\/{\rm m}^{-3}$~\citep{patzold2019nucleus}.
We first consider $\tstar$ (blue line) for an orbit with eccentricity $e=0.1$,
From \subfig{fig:param}{B}, we know that for such an orbit
erosion happens if $R \lesssim 9{\rm au}$.
Hence, we plot $\tstar$ for $R \lesssim 9{\rm au}$. 
For orbits with $R$ less than this limit $\tstar$ ranges
from less than a year to about $200$ years, extremely short times
in astronomical time scales.
For orbits with $R$ larger than this value erosion has no effect. 
For an orbit with eccentricity $e = 0.05$, $\tstar$
[orange line in \subfig{fig:param}{C}] ranges from less than a year
to about $300$ years, only if $R \lesssim 8 {\rm au}$.
Even for a circular orbit (green line) $\tstar$ ranges from about $100$ years
to little less than ten thousand years, only if $R \lesssim 0.6{\rm au}$ .
Hence, we conclude that planetesimals in eccentric orbits, of even very
small eccentricity, rapidly (in about hundred years) erodes away if the
semi-major axis of their
orbit lies in the inner disk (less than about $10$ au). 
Even planetesimals in circular orbits erode away in about ten thousand
years if the semi-major axis of their orbits are closer than $0.9{\rm au}$.

\section{Conclusion}
\label{sec:conc}
In a recent paper \cite{rozner2020aeolian}, have argued that under erosion
 $dL/dt \sim 1/L$.
This is different from the law, \Eq{eq:mloss4}, we report.
Our result is supported by theory~\citep{ristroph2012sculpting,moore2013self},
our numerical simulations and experiments~\citep{ristroph2012sculpting}. 
Furthermore, unlike us, \cite{rozner2020aeolian} do not take into account the
dynamics of the problem, i.e., the fact that the fluid stress eroding the
body changes as the body erodes.
However, they also reach the same
qualitative conclusion that erosion is rapid.

Let us repeat that we find, contrary to our earlier
work~\citep{schaffer2020erosion}, that 
erosion happens even for boulders in perfectly circular orbits
if they are close enough to the central star.
This is so because of two reasons: (a) in the light of recent
  experimental results~\citep{demirci2020planetesimals}, we consider
  a lower value of $\taus$ and 
  (b) we consider the typical fluid stress
  not the minimum value as we had done before.

 \subsection{How robust are our results?}
The law of erosion, \Eq{eq:mloss4}, is derived under several
simplifying assumptions. 
Our simulations, which are not limited by those
assumptions, for the first time, show its universal nature
-- the law holds for all the shapes and the Reynolds numbers 
we consider, irrespective of whether the laminar
boundary layer has become unstable or not. 
But the expression for the time it takes for the body to erode
away, $\tstar$, does not follow the simple theory. 
At small Reynolds numbers, it is typically larger than
the theoretical prediction, at large Reynolds numbers it is typically 
smaller. 
We estimate the Reynolds number of a $10$km sized boulder to be
$10^4$ or larger, \Fig{fig:param}. 
Hence, we expect that in reality the time it takes for
a $10$km sized planetesimal to erode away is 
shorter than the $\tstar$ we estimate in
\subfig{fig:param}{C}. 

The estimate of the dimensionless threshold ($\Th$) is less
certain.  Our estimate for the threshold stress, $\taus$,
may be a gross underestimate 
if the planetesimal contains snow. 
Hence, we expect effects of erosion to be small beyond
the snow line. 
Furthermore, different layers on the planetesimal may
have different threshold stress; the inner layers
may be more strongly held due to sintering. 
Once exposed, it may take longer to erode them. 
It is straightforward to add such effects to our simulations
but is futile as we do not know quantitatively the
effects of sintering in planetesimals. 
Note than even if $\taus$ increases by a factor of $10$ or
$100$, erosion will still occur although at shorter distance from the
central star or in orbits with higher eccentricity. 

The estimate of the erosion number is also uncertain
because of the uncertainty regarding $\ker$.
We know of only one experiment~\citep{demirci2020planetesimals} from which we estimate $\ker$.
If this number is smaller by a factor of $10$, $\tstar$
increases by a factor of $10$. 
Even then erosion is rapid in astronomical time scales. 

Naturally, erosion is also accompanied by deposition.
Deposition also happens with a threshold stress but this threshold
is typically lower than the erosion threshold~\citep{salles1993deposition}.
In the range between these two thresholds the body neither grows nor
decays.
In this paper we have ignored deposition.
It is possible that planetesimals close to the central star erode,
while the material that is eroded is deposited on the planetesimals
further away such that their growth rate actually increases. 

Finally, note that in a protoplanetary disk the gas flow is turbulent,
whereas in our model (both theoretical and numerical) we have assumed the
incoming flow to be laminar.
Very little is known about drag, lift, or wall stress of
bodies in flows that are already turbulent. 
We can speculate that in such cases we will have an even
thinner and highly fluctuating boundary layer.
This we leave for the future. 

\subsection{Application to objects in the asteroid belt}
  How do we reconcile our results with the fact that the
asteroid belt of the solar system has many objects with sizes ranging from
about a kilometer to hundreds of kilometers  in eccentric orbits?
The asteroid belt lies between two and three au.
As a specific example, consider the minor planet Vesta, which is about $500$ km
in size in an orbit with eccentricity about $0.09$ and a semimajor
axis of about  $2.3$au.
According to our theory such asteroids could not form
where they are at present by mere aggregation because they would have
eroded away as soon as they formed. 
This gives rise to several possibilities. 
One, Vesta was formed originally on a circular orbit but developed the small
eccentricity it now has at a later stage when the gas in the disk had
disappeared.
Two, Vesta originally formed further away 
in the disk and had migrated inward at a later stage. 
Three -- the most interesting one -- these asteroids originally
formed, by gravitational collapse, as
much bigger bodies and have eroded away to their present size 
in a time scale of about a megayear -- the
approximate lifetime of the disk. We discuss this
possibility next. 

Consider the possibility that the gravitational collapse
creates a body of approximate size of \ATT{$10^3$} km. 
To apply our results to such a body  we must also include the gravitational pull by the body itself.
Figure 7 in the article by \citep{demirci2020planetesimals}
suggests that the threshold stress increases linearly with gravity
with a proportionality constant
$\alpha \approx 7 \times 10^{-2}$ in units of
kilograms divided by meter squared.
Hence, the threshold stress of a $10^3$ km body is
$\taus\approx   0.001 {\rm Pa} + \alpha g$
where $g$ is the gravitational acceleration on the surface
of the asteroid given by: $g =  g_{\earth} L/R_{\earth}$
where $g_{\earth} \approx 9.8 {\rm ms}^{-2}$ is the
acceleration on the surface of Earth and
$R_{\earth} \approx 6400$ km is the radius of Earth. 
For a body of size approximately $10^{3}$ km
we obtain $\taus \approx 0.1$ Pa.  
For this case, our calculations show that the dimensionless
threshold is greater than unity for an orbit with $e=0.1$ up to
a distance of about $2$ au.
We further find that such a body will erode away in about $10^5$
years. However, the value of the constant $\ker$, which determines
the rate of erosion, is not known accurately and also, like $\taus$,  should
depend on gravity.  We have no experimental data on this.
Clearly, larger gravity implies that $\ker$ is smaller.
If we consider a much smaller $\ker \approx 8{\rm m}^{-1}{\rm s}$
we find that  $\tstar \approx 8\times 10^{6}$ years.
In other words, if we consider the lifetime of the disk to be about $10^6$ years
a $10^3$ km body at a distance of about $2$ au that has formed by
gravitational collapse on a orbit with eccentricity of $0.1$ will erode
away \textit{partially but not completely}.
After being eroded for $10^6$ years the size of the body is going to be
approximately $300$ km. Of course, none of the specific
numbers in this paragraph are supposed to be precise.
Thus, we illustrate that our theory is consistent with the recently
suggested hypothesis~\citep{klahr2020turbulence, klahr2021testing}
 that the planetesimals form by gravitational
collapse to bodies of about $100$~km in size or larger.

\section*{Acknowledgements}
The code to plot \Fig{fig:snowman} was written by Aritra Bhakat. All figures in
this paper are plotted using the free software matplotlib~\citep{Hunter:2007}.
DM thanks Srikanth Toppaladoddi, Alessandro Morbideli, and John Wettlaufer for
stimulating discussions.
We thank Prasad Perlekar for helping us write the lattice Boltzmann code.

This work is partially  funded  by  the  ``Bottlenecks  for particle  growth
in  turbulent  aerosols''  grant  from  the  Knut  and  Alice  Wallenberg  Foundation  (2014.0048).  
In addition, A.J. acknowledges funding from the 
Swedish  Research  Council  (grant 2014-5775),  the  Knut  and  
Alice  Wallenberg  Foundation  (grants 2012.0150,2014.0017) and the 
European Research Council (ERC ConsolidatorGrant724687-PLANETESYS) for research support.
B.M. acknowledges funding from the 
Swedish Research Councils (grant 2017-3865)
D.M. acknowledges funding from the Swedish Research Council ( 638-2013-9243, 2016-05225). The simulations were performed on resources provided by the Swedish National Infrastructure for Computing (SNIC) at PDC center for High Performance
Computing. 
\appendix
\section{Lattice Boltzmann Method}
\label{sec:LBM}
Instead of solving the Navier--Stokes equation the Lattice Boltzmann
Method solves the Boltzmann equation on a Cartesian lattice.
Recall, that the Boltzmann equation is an equation of evolution of
probability density function, $f(\xx,\vv,t)$, of molecules in
\textit{phase space},
where $\xx$ is the physical coordinates and $\vv$ is the velocity coordinates
of phase space~\cite[see, e.g.,][]{LLphykin}. 
A hydrodynamic description of the system emerges on averaging over the
phase-space, i.e, 
the hydrodynamic density,
\begin{equation}
  \rho(\xx,t) \equiv \int f(\xx,\vv,t) d^dv\/,
  \label{eq:rho}
\end{equation}
and the hydrodyamic momentum,
\begin{equation}
  \rho\uu(\xx,t) \equiv \int \vv f(\xx,\vv,t) d^d v \/,
  \label{eq:mom}
\end{equation}
The  hydrodynamic pressure and the stress tensor
emerges as respectively the isotropic and non-isotropic part of  
\begin{equation}
  \sigma_{\ab} \equiv \int \cab f(\xx,\vv,t)d^d v\/,
  \label{eq:stress}
\end{equation}
where the Greek indices denote Cartesian components and
$\ca \equiv \vv - \uu$.
Thus, once we have numerically solved the Boltzmann equation
it is straightforward to obtain the hydrodynamic variables, which are
guaranteed to satisfy the Navier-Stokes equation.
The proof of this last statement is through the Chapman--Enskog
expansion. The proof becomes significantly simpler if
the collision integral on the right hand side of the Boltzmann
equation is replaced by its Bhatnagar--Gross--Krook~(BGK) approximation, 
which postulates that the only effect of collision is that
at every grid point in physical space $f$ relaxes to
its equilibrium value -- a Maxwellian distribution -- with a
single characteristic time-scale $\tau$.  Within the BGK approximation,
the kinematic viscosity of the fluid is
\begin{equation}
  \nu = (\tau -1/2)
\label{eq:nu}
\end{equation}
To solve the Boltzmann equation numerically we need to discretize the
physical space but more importantly also the velocity space.
The velocity space is discretized into 27 discrete
lattice vectors, this is known as the D3Q27 model of the LBM.
These lattice vectors are plotted in \Fig{fig:fi3d}.
Once we solve for $f$ by solving the discrete Boltzmann equation
it is straightforward to calculate the velocity, density, and the stress tensors
from $f$ by replacing the integral in equations \ref{eq:rho}, \ref{eq:mom}, and
\ref{eq:stress} by a sum over the twenty seven discrete values of velocity.

A major advantage of this method is the way it deals with boundary conditions.
We use a technique called \textit{bounce back} to model no-slip boundary
conditions,
as we show in \Fig{fig:bback}.
The shaded part of the figure is the solid and the grid points there are
classified as solid grid points.
The physical boundary is imagined halfway between the grid points,
the boundary between the shaded and the unshaded region.
The $\vv\cdot\grad f$ term
in the Boltzmann equation denotes streaming of the component of
$f$ along a particular
lattice vector $\qq$ by the velocity along that direction.
In \subfig{fig:bback}{a}
show a grid point with three lattice vectors.
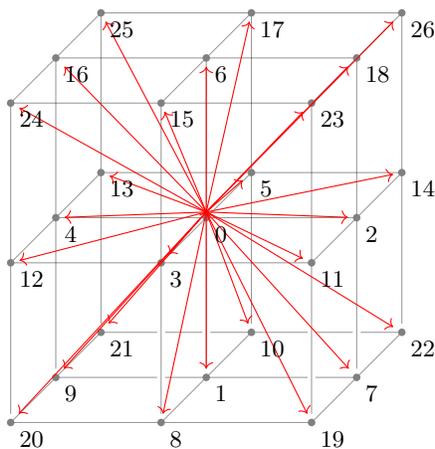
\begin{figure}[h!]
\centering
\begin{tikzpicture}
    \begin{scope}
        \myGlobalTransformation{0}{0};
        \graphLinesHorizontal;
        \graphLinesVertical;

        \foreach \x in {1,3,5} {
            \foreach \y in {1,3,5} {
                \node (thisNode) at (\x,\y) {};
                {
                    \pgftransformreset
                    \draw[white, myBG] (thisNode) -- ++(0,4.25);
                    \draw[black,opacity = 0.4] (thisNode) -- ++(0,4.25);
                }
            }
        }
    \end{scope}
    
    \begin{scope}
        \myGlobalTransformation{0}{2.125};
        \graphLinesHorizontal;
        \graphLinesVertical;
    \end{scope}

    \begin{scope}
        \myGlobalTransformation{0}{4.25};
        \graphLinesHorizontal;
        \graphLinesVertical;
    \end{scope}

    \graphThreeDnodes{0}{0};
    \graphThreeDnodes{0}{4.25}; 
    \graphThreeDnodes{0}{2.125};
    
    \farrows{0}{0};
    \farrows{0}{2.125};
    \farrows{0}{4.25};
    
    \numberNodes{0}{0};
\end{tikzpicture}
\caption{3D visualization of the twenty seven discrete lattice vectors (one of them is zero) used to discretize the
  distribution function $f$ in lattice Boltzmann algorithm D3Q27. }
    \label{fig:fi3d}
\end{figure}
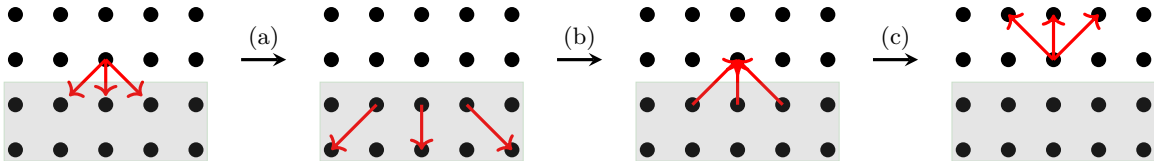
\begin{figure*}[h!]

    \begin{tikzpicture}[scale = 0.3]
        
    \begin{scope}
        \foreach \x in {-1,1,3,5,7} {
            \foreach \y in {-1,1,3,5} {
                \node at (\x,\y) [circle,fill=black,minimum size = 0.2cm, inner sep = 0cm] {};
            }
        }
        \foreach \x in {1,3,5} {
            \foreach \y in {1} {
                \node (thisNode) at (\x,\y) {};
                {
                    \draw[->, red, very thick] (3 , 3) -- (thisNode) node[above right] {};
                }
            }
        }
        \filldraw[fill=gray, draw=green!40!black, opacity=0.2] (-1.5,-1.5) rectangle (7.5,2);
    \end{scope}
    
    
    \draw[->, black, very thick, -stealth] (9,3) -- (11,3) node[midway, above] {(a)};
    
    \begin{scope}
        \foreach \x in {13,15,17,19,21} {
            \foreach \y in {-1,1,3,5} {
                \node at (\x,\y) [circle,fill=black,minimum size = 0.2cm, inner sep = 0cm] {};
            }
        }
        \draw[->, red, very thick] (15,1) -- (13,-1) node[above right] {}; 
        \draw[->, red, very thick] (17,1) -- (17,-1) node[above right] {}; 
        \draw[->, red, very thick] (19,1) -- (21,-1) node[above right] {}; 
        \filldraw[fill=gray, draw=green!40!black, opacity=0.2] (12.5,-1.5) rectangle (21.5,2);
    \end{scope}
        
    \draw[->, black, very thick, -stealth] (23,3) -- (25,3) node[midway, above] {(b)};
        
    \begin{scope}
        \foreach \x in {27,29,31,33,35} {
            \foreach \y in {-1,1,3,5} {
                \node at (\x,\y) [circle,fill=black,minimum size = 0.2cm, inner sep = 0cm] {};
            }
        }
        \draw[->, red, very thick] (29,1) -- (31,3) node[above right] {}; 
        \draw[->, red, very thick] (31,1) -- (31,3) node[above right] {}; 
        \draw[->, red, very thick] (33,1) -- (31,3) node[above right] {}; 
        \filldraw[fill=gray, draw=green!40!black, opacity=0.2] (26.5,-1.5) rectangle (35.5,2);
    \end{scope}
    \draw[->, black, very thick, -stealth] (37,3) -- (39,3) node[midway, above] {(c)};
    \begin{scope}
        \foreach \x in {41,43,45,47,49} {
            \foreach \y in {-1,1,3,5} {
                \node at (\x,\y) [circle,fill=black,minimum size = 0.2cm, inner sep = 0cm] {};
            }
        }
        \draw[->, red, very thick] (45,3) -- (43,5) node[above right] {}; 
        \draw[->, red, very thick] (45,3) -- (45,5) node[above right] {}; 
        \draw[->, red, very thick] (45,3) -- (47,5) node[above right] {}; 
        \filldraw[fill=gray, draw=green!40!black, opacity=0.2] (40.5,-1.5) rectangle (49.5,2);
    \end{scope}
    \end{tikzpicture}
    \caption{Bounce back sequence. (a) is the streaming step. (b) is the
      bounce back boundary condition. (c) is streaming step again.
      Gray area represents
      the solid domain while the white represents the fluid domain.}
    \label{fig:bback}
\end{figure*}
\subsection{Implementation of erosion}
\label{sec:dns_erosion}
Our numerical scheme follows \cite{jager2017channelization}.
Let the deviatoric stress tensor be
\begin{equation}
  \sigma_{\ab} \equiv \mu \left(\dela u_{\beta} + \delb u_{\alpha} \right)\/.
  \label{eq:stress2}
\end{equation}
The shear force at an infinitesimal surface element $dS$ with unit normal $\nhat$
is given by $F_{\alpha} = \sigma_{\alpha\beta}n_{\beta}$.
The magnitude of the tangential
component of this force is the wall shear stress
\begin{equation}
  \tauf = \mid \FF - \nhat\cdot\FF  \mid
  \label{eq:tauf2}
\end{equation}
In the lattice Boltzmann method this is calculated as
\begin{equation}
\sigma_{\ab} = \Bigg(1-\frac{1}{2 \tau} \Bigg) \sum_i f^{\rm{neq}}_{i}(\mathbf{c_i})^a (\mathbf{c_i})^b,
\end{equation}
where $f^{\rm{neq}}_{i}=f_{i}-f^{\rm{eq}}_{i}$
is the nonequilibrium part of the distribution
function~\citep{jager2017channelization}
and $i$ runs over the lattice vectors -- $27$ in the D3Q27 model.

\bibliographystyle{aasjournal}

%

%
\end{document}